\documentclass[12pt]{article}
\begin{document}
\def\bfone{\relax{\rm 1\kern-.35em 1}}\let\shat=\hat
\def\hat{\widehat}

%
\def\cS{{\cal K}}
\def\IE{\relax{{\rm I\kern-.18em E}}}
\def\cE{{\cal E}}
\def\rt{{\cR^{(3)}}}
\def\IGam{\relax{{\rm I}\kern-.18em \Gamma}}
\def\IGa{\IA}
\def\cV{{\cal V}}
\def\Rt{{\cal R}^{(3)}}
\def\tft#1{\langle\langle\,#1\,\rangle\rangle}
\def\IA{\relax{\hbox{{\rm A}\kern-.82em {\rm A}}}}
\def\hata{{\shat\a}}
\def\hatb{{\shat\b}}
\def\hatA{{\shat A}}
\def\hatB{{\shat B}}
\def\bv{{\bf V}}
\def\Fb{\overline{F}}
\def\nablab{\overline{\nabla}}
\def\Ub{\overline{U}}
\def\Db{\overline{D}}
\def\zb{\overline{z}}
\def\eb{\overline{e}}
\def\fb{\overline{f}}
\def\tb{\overline{t}}
\def\Xb{\overline{X}}
\def\Vb{\overline{V}}
\def\Cb{\overline{C}}
\def\Sb{\overline{S}}
\def\delb{\overline{\del}}
\def\Gammab{\overline{\Gamma}}
\def\Ab{\overline{A}}
\def\Anh{A^{\rm nh}}
\def\alphab{\bar{\alpha}}
\def\cy{Calabi--Yau}
\def\cabg{C_{\alpha\beta\gamma}}
\def\B{\Sigma}
\def\Bh{\hat \Sigma}
\def\Kh{\hat{K}}
\def\Knh{{\cal K}}
\def\A{\Lambda}
\def\Ah{\hat \Lambda}
\def\R{\hat{R}}
\def\V{{V}}
\def\T{T}
\def\Gammah{\hat{\Gamma}}
\def\twot{$(2,2)$}
\def\K{K\"ahler}
\def\rat{({\theta_2 \over \theta_1})}
\def\lv{{\bf \omega}}
\def\w{w}
\def\CP{C\!P}
\def\o#1#2{{{#1}\over{#2}}}
\def\eq#1{(\ref{#1})}

\def\ib{{\bar \imath}}
\def\jb{{\bar \jmath}}
\def\Im{{\rm Im ~}}
\def\Re{{\rm Re ~}}
\def\IP{\relax{\rm I\kern-.18em P}}
\def\arccosh{{\rm arccosh ~}}
%
\font\cmss=cmss10 \font\cmsss=cmss10 at 7pt
\def\twomat#1#2#3#4{\left(\matrix{#1 & #2 \cr #3 & #4}\right)}
\def\inbar{\vrule height1.5ex width.4pt depth0pt}
\def\IC{\relax\,\hbox{$\inbar\kern-.3em{\rm C}$}}
\def\IG{\relax\,\hbox{$\inbar\kern-.3em{\rm G}$}}
\def\IB{\relax{\rm I\kern-.18em B}}
\def\ID{\relax{\rm I\kern-.18em D}}
\def\IL{\relax{\rm I\kern-.18em L}}
\def\IF{\relax{\rm I\kern-.18em F}}
\def\IH{\relax{\rm I\kern-.18em H}}
\def\II{\relax{\rm I\kern-.17em I}}
\def\IN{\relax{\rm I\kern-.18em N}}
\def\IP{\relax{\rm I\kern-.18em P}}
\def\IQ{\relax\,\hbox{$\inbar\kern-.3em{\rm Q}$}}
\def\bfzero{\relax\,\hbox{$\inbar\kern-.3em{\rm 0}$}}
\def\IR{\relax{\rm I\kern-.18em R}}
\def\ZZ{\relax\ifmmode\mathchoice
{\hbox{\cmss Z\kern-.4em Z}}{\hbox{\cmss Z\kern-.4em Z}}
{\lower.9pt\hbox{\cmsss Z\kern-.4em Z}}
{\lower1.2pt\hbox{\cmsss Z\kern-.4em Z}}\else{\cmss Z\kern-.4em
Z}\fi}
\def\IU{\relax\,\hbox{$\inbar\kern-.3em{\rm U}$}}
\def\bfone{\relax{\rm 1\kern-.35em 1}}
\def\dop{{\rm d}\hskip -1pt}
\def\real{{\rm Re}\hskip 1pt}
\def\trace{{\rm Tr}\hskip 1pt}
\def\ii{{\rm i}}
\def\diag{{\rm diag}}
\def\sch#1#2{\{#1;#2\}}

\newcommand{\ft}[2]{{\textstyle\frac{#1}{#2}}}
\newcommand{\QED}{{\hspace*{\fill}\rule{2mm}{2mm}\linebreak}}
\def\dop{{\rm d}\hskip -1pt}
\def\bfone{\relax{\rm 1\kern-.35em 1}}
\def\bfzero{\relax{\rm I\kern-.18em 0}}
\def\inbar{\vrule height1.5ex width.4pt depth0pt}
\def\IC{\relax\,\hbox{$\inbar\kern-.3em{\rm C}$}}
\def\ID{\relax{\rm I\kern-.18em D}}
\def\IF{\relax{\rm I\kern-.18em F}}
\def\IK{\relax{\rm I\kern-.18em K}}
\def\IH{\relax{\rm I\kern-.18em H}}
\def\II{\relax{\rm I\kern-.17em I}}
\def\IN{\relax{\rm I\kern-.18em N}}
\def\IP{\relax{\rm I\kern-.18em P}}
\def\IQ{\relax\,\hbox{$\inbar\kern-.3em{\rm Q}$}}
\def\IR{\relax{\rm I\kern-.18em R}}
\def\IG{\relax\,\hbox{$\inbar\kern-.3em{\rm G}$}}
\font\cmss=cmss10 \font\cmsss=cmss10 at 7pt
\def\ZZ{\relax\ifmmode\mathchoice
{\hbox{\cmss Z\kern-.4em Z}}{\hbox{\cmss Z\kern-.4em Z}}
{\lower.9pt\hbox{\cmsss Z\kern-.4em Z}}
{\lower1.2pt\hbox{\cmsss Z\kern-.4em Z}}\else{\cmss Z\kern-.4em
Z}\fi}
\def\i{\rm i} 
\def\a{\alpha} \def\b{\beta} \def\d{\delta}
\def\e{\epsilon} \def\c{\gamma}
\def\G{\Gamma} \def\l{\lambda}
\def\L{\Lambda} \def\s{\sigma}
\def\cA{{\cal A}} \def\cB{{\cal B}}
\def\cC{{\cal C}} \def\cD{{\cal D}}
\def\cF{{\cal F}} \def\cG{{\cal G}}
\def\cH{{\cal H}} \def\cI{{\cal I}}
\def\cJ{{\cal J}} \def\cK{{\cal K}}
\def\cL{{\cal L}} \def\cM{{\cal M}}
\def\cN{{\cal N}} \def\cO{{\cal O}}
\def\cP{{\cal P}} \def\cQ{{\cal Q}}
\def\cR{{\cal R}} \def\cV{{\cal V}}\def\cW{{\cal W}}
%
%
%
\def\crr{\crcr\noalign{\vskip {8.3333pt}}}
\def\tilde{\widetilde}
\def\bar{\overline}
\def\us#1{\underline{#1}}
\let\shat=\hat
\def\hat{\widehat}
\def\hyp{\vrule height 2.3pt width 2.5pt depth -1.5pt}
\def\square{\mbox{.08}{.08}}

\def\Coeff#1#2{{#1\over #2}}
\def\Coe#1.#2.{{#1\over #2}}
\def\coeff#1#2{\relax{\textstyle {#1 \over #2}}\displaystyle}
\def\coe#1.#2.{\relax{\textstyle {#1 \over #2}}\displaystyle}
\def\half{{1 \over 2}}
\def\shalf{\relax{\textstyle {1 \over 2}}\displaystyle}
\def\dag#1{#1\!\!\!/\,\,\,}
\def\to{\rightarrow}
\def\notin{\hbox{{$\in$}\kern-.51em\hbox{/}}}
\def\shdot{\!\cdot\!}
\def\ket#1{\,\big|\,#1\,\big>\,}
\def\bra#1{\,\big<\,#1\,\big|\,}
\def\equaltop#1{\mathrel{\mathop=^{#1}}}
\def\Trbel#1{\mathop{{\rm Tr}}_{#1}}
\def\inserteq#1{\noalign{\vskip-.2truecm\hbox{#1\hfil}
\vskip-.2cm}}
\def\attac#1{\Bigl\vert
{\phantom{X}\atop{{\rm\scriptstyle #1}}\phantom{X}}}
\def\exx#1{e^{{\displaystyle #1}}}
\def\del{\partial}
\def\delbar{\bar\partial}
\def\nex#1{$N\!=\!#1$}
\def\dex#1{$d\!=\!#1$}
\def\cex#1{$c\!=\!#1$}
\def\eg{{\it e.g.}} \def\ie{{\it i.e.}}
%
\def\cS{{\cal K}}
\def\IE{\relax{{\rm I\kern-.18em E}}}
\def\cE{{\cal E}}
\def\rt{{\cR^{(3)}}}
\def\IGam{\relax{{\rm I}\kern-.18em \Gamma}}
\def\IGa{\IA}
\def\cV{{\cal V}}
\def\Rt{{\cal R}^{(3)}}
\def\tft#1{\langle\langle\,#1\,\rangle\rangle}
\def\IA{\relax{\hbox{{\rm A}\kern-.82em {\rm A}}}}
\def\hata{{\shat\a}}
\def\hatb{{\shat\b}}
\def\hatA{{\shat A}}
\def\hatB{{\shat B}}
\def\bv{{\bf V}}
\def\Fb{\overline{F}}
\def\nablab{\overline{\nabla}}
\def\Ub{\overline{U}}
\def\Db{\overline{D}}
\def\zb{\overline{z}}
\def\eb{\overline{e}}
\def\fb{\overline{f}}
\def\tb{\overline{t}}
\def\Xb{\overline{X}}
\def\Vb{\overline{V}}
\def\Cb{\overline{C}}
\def\Sb{\overline{S}}
\def\delb{\overline{\del}}
\def\Gammab{\overline{\Gamma}}
\def\Ab{\overline{A}}
\def\Anh{A^{\rm nh}}
\def\alphab{\bar{\alpha}}
\def\cy{Calabi--Yau}
\def\cabg{C_{\alpha\beta\gamma}}
\def\B{\Sigma}
\def\Bh{\hat \Sigma}
\def\Kh{\hat{K}}
\def\Knh{{\cal K}}
\def\A{\Lambda}
\def\Ah{\hat \Lambda}
\def\R{\hat{R}}
\def\V{{V}}
\def\T{T}
\def\Gammah{\hat{\Gamma}}
\def\twot{$(2,2)$}
\def\K{K\"ahler}
\def\rat{({\theta_2 \over \theta_1})}
\def\lv{{\bf \omega}}
\def\w{w}
\def\CP{C\!P}
\def\o#1#2{{{#1}\over{#2}}}
\def\eq#1{(\ref{#1})}
\newcommand{\be}{\begin{equation}}
\newcommand{\ee}{\end{equation}}
\newcommand{\ba}{\begin{eqnarray}}
\newcommand{\ea}{\end{eqnarray}}
\newtheorem{definizione}{Definition}[section]
\newcommand{\bd}{\begin{definizione}}
\newcommand{\ed}{\end{definizione}}
\newtheorem{teorema}{Theorem}[section]
\newcommand{\bth}{\begin{teorema}}
\newcommand{\eth}{\end{teorema}}
\newtheorem{lemma}{Lemma}[section]
\newcommand{\blem}{\begin{lemma}}
\newcommand{\elem}{\end{lemma}}
\newcommand{\brr}{\begin{array}}
\newcommand{\err}{\end{array}}
\newcommand{\nn}{\nonumber}
\newtheorem{corollario}{Corollary}[section]
\newcommand{\bcorol}{\begin{corollario}}
\newcommand{\ecorol}{\end{corollario}}
\def\twomat#1#2#3#4{\left(\begin{array}{cc}
 {#1}&{#2}\\ {#3}&{#4}\\
\end{array}
\right)}
\def\twovec#1#2{\left(\begin{array}{c}
{#1}\\ {#2}\\
\end{array}
\right)}
\thispagestyle{empty}
\begin{titlepage}
\thispagestyle{empty}
\begin{flushright}
CERN-TH/98-91\\
hep-th/9803171\\
\end{flushright}
\vskip 2.cm
\begin{center}
{\Large\bf
 K--K excitations on $AdS_5\times S^5$ as  $N=4$ ``primary'' superfields.
\renewcommand{\thefootnote}{\fnsymbol{footnote}}\footnote{Work
 supported in part by EEC under TMR contract ERBFMRX-CT96-0045 (LNF Frascati,
 Politecnico di Torino), Angelo Della Riccia fellowship and by DOE grant
 DE-FG03-91ER40662
 }}
\vskip 2.cm
{\large L. Andrianopoli$^{a,b}$ and S. Ferrara$^a$}
\end{center}

{\it $^a$ CERN Theoretical Division, CH 1211 Geneva 23, Switzerland.}

{\it $^b$  Istituto Nazionale di Fisica Nucleare (INFN)-Sezione di
 Torino, Italy.} 
\begin{abstract}
We show that the K--K spectrum of IIB string on $AdS_5 \times S_5$ is described
by ``twisted chiral'' $N=4$ superfields, naturally described in ``harmonic superspace'', obtained by taking suitable gauge
 singlets polynomials of the $D$3-brane boundary $SU(n)$ superconformal field
 theory.

To each $p$-order polynomial is associated a massive K--K short representation
 with 
$256 \times \frac{1}{12} p^2(p^2 -1) $ states.
The $p=2$ quadratic polynomial corresponds to the ``supercurrent multiplet'' 
describing the ``massless'' bulk graviton multiplet.
\end{abstract}
\vskip 2cm
\begin{flushleft}
  CERN-TH/98-91\\
  March, 1998
\end{flushleft}
\end{titlepage}
\section{Introduction}

In recent time increasing evidence of a close connection between $AdS_{p+2}$
 supergravities
and $p+1$ dimensional superconformal field theories has emerged \cite{mal}--\cite{dlp}.

The original duality between these theories was proposed in the context where
 the latter describe the dynamics of the degrees of freedom of $p$-branes and
 the former the near horizon geometry \cite{kleb}.
However it is appealing to investigate whether this correspondence is more
 general and valid in more general contexts.

Indeed the proposal of relating the correlators of superconformal field 
theories to the classical Anti de Sitter supergravity partition function \cite{gkp},\cite{wit} may
 be a suitable tool to investigate non-perturbative properties of scale
 invariant field theories \cite{wit2}.
In a more geometrical context these proposals are a realisation of the 
previous attempt \cite{flfr} to describe ``elementary massless particles'' on AdS in terms
 of ``more elementary constituents'', the singletons \cite{fronsdal}.

Indeed it was recently shown \cite{ff1},\cite{ff2},\cite{gkp},\cite{wit},\cite{ffz}
 that massless particles, associated to 
diffeomorphisms, supersymmetry and Yang--Mills symmetry in the AdS bulk,
 correspond to particular ``conserved
currents'' composite operators in the boundary singleton theory where $\del
 AdS_{p+2} =
 \tilde \cM_{p+1}$ and $\tilde \cM_{p+1} \sim S^p \times R $ is a particular
 completion of  Minkowski space.
In particular it was shown that some of the K--K excitations of ten
 dimensional type IIB
supergravity on $AdS_5 \times S^5 $ \cite{grw2}
 can be described taking suitable 
polynomials of $N=1,2$  chiral multiplets where the $N=4$ theory is analyzed 
in terms of superfields of lower supersymmetry \cite{wit}.

It was then realized that these K--K excitations  are contained in 
$N=4$ superfield towers where apparent different towers do indeed correspond to
 different $\theta$ components of the same superfield \cite{ffz}.

It is the aim of this note to complete this analysis, by using the description
 of $N=4$ superfields through the use of extra suitable variables $u^A_r$
 ($r=1,2$ $A=1,\cdots ,4$) which are
 coordinates of the coset 
$\frac{SU(4)}{S(U(2) \times U(2))}$ \cite{gikos},\cite{hw}.

In particular we will show that the analysis in \cite{ffz} does indeed give the
 entire K--K spectrum of
massless and massive states of IIB string compactified on $AdS_5 \times S^5$
\cite{grw2}.

The $N=4$ superfields obtained by taking the singlet of the $p$-order
 polynomial of the $N=4$ 
gauge singleton superfield, in the $p$-symmetric traceless representation of
 $SU(4)$ (corresponding
to Dynkin labels $(0,p,0)$ \cite{slan}), does indeed give a short representation containing
 $256 \times \frac{1}{12}
p^2(p^2-1)$ states whose highest spin 2 state is in the $p-2$ symmetric
 traceless representation of $SU(4)$.

\section{The K--K spectrum of IIB on $AdS_5 \times S^5$}

The K--K spectrum of maximal supergravity on $AdS_5 \times S^5$ \cite{grw2}
 was given, in
 terms of unitary irreducible
representations (U.I.R.) of the superalgebra $SU(2,2 / 4)$, in \cite{gm}.
The masses of the individual spin states were given in \cite{krv}.

Let us briefly recall that the singleton representation corresponds to the
 gauge-field multiplet on the boundary.
It is the representation with $p=1$ of \cite{gm}.
To correctly count the states one must take into account the gauge modes on
 the boundary which
must be subtracted from the original spin assignments.
Indeed the singleton representation corresponds to a ``massless'' 
representation of the same superalgebra
acting as superconformal algebra on the four dimensional boundary of $AdS_5$.

Therefore the singleton multiplet contains $2^4$ states, eight bosons with 
helicities $\pm 2$ ({\bf 1}) and zero ({\bf 6})
and eight fermions with helicities $+\half$ ({\bf 4}) and $-\half$
 ($\bar{\bf 4}$) where
 in brackets we denoted $SU(4)$ representations.

The massless representation on $AdS_5$, which corresponds to the graviton
 multiplet of $AdS_5$, is 
given by the $p=2$ towers of \cite{gm}.
Again, due to gauge modes in the bulk, to correctly count the states one must
 subtract them from
the spin 2, $\frac{3}{2}$ and 1 gauge fields.

The gauge modes in the bulk, which must be subtracted from the $(1,1)$, 
$(\half , 1) + (1, \half)$ and $(\half , \half)$ $O(4)$ reps. describing 
the gauge fields, correspond to $(\half , \half)$, $(\half , 0) +
 (0, \half)$ and $(0,0)$ $O(4)$ representations. 

When this is done, the massless multiplet is seen to contain precisely 256 
states, 128 bosons and 128 fermions.

These states are given in ref. \cite{gm} with the proviso that, due to the 
gauge modes, the three above reps.
should account for five, eight and three states respectively (instead of nine,
 twelve and four).

The massive K--K towers are given from the table in ref. \cite{gm} for $p\ge 3$.
For massive states no gauge modes are present so states are classified by 
$(E_0, J_1, J_2)$ labels with 
multiplicity $(2J_1 + 1)(2J_2 + 1)$ of a given $O(4)$ representation.

A $SU(4)$ representation, with Dynkin labels $(a_1, a_2, a_3)$ \cite{slan},
 has dimension \footnote{We thank Leonardo Castellani for a useful discussion
on this point.}
\begin{eqnarray}
  d(a_1, a_2, a_3)=\hfill \nn\\
(a_1+1)(a_2+1)(a_3+1)\left(1+ \frac{a_1 + a_2}{2}\right)
\left(1+ \frac{a_2 + a_3}{2}\right)\left(1+ \frac{a_1 + a_2 + a_3}{3}\right)
\end{eqnarray}

All the states with a given $p$ must be included.
This gives $256 \times \frac{1}{12} p^2(p^2-1)$ states, half bosons and half
 fermions. 

Note in particular that such representation, uniquely denoted by $p$, contains
 a scalar state with conformal
weight $E_0 =p$ in the $(0,p,0)$ $SU(4)$ rep., a spin 2 (massive graviton)
 state with middle weight
$E_0 =p+2$ in the $(0,p-2,0)$ rep. of $SU(4)$ and a (real) scalar in the 
$(0,p-4,0)$ rep. of $SU(4)$
with maximal weight $E_0 = p+4$ ($p\ge 4$).

It also appears from the table of \cite{gm} that the representations for
 $p=1,2,3$ are not generic,
 while for $p\ge 4$ they are generic. 
We will understand in the next section why this is the case.

\section{``Twisted chiral superfields'' and K--K states as composite singleton
 excitations}

In the present section we would like to complete the proof of the assertion of
 ref. \cite{ffz} that the 
entire K--K spectrum of IIB on $AdS_5 \times S^5$ is given by a single tower 
of $SU(n)$ singlets obtained by taking the traces
of a $p$-order polynomial of a $SU(n)$ Lie algebra valued $N=4$ singleton 
superfield, whose $\theta =0$ term 
corresponds to the $(0,p,0)$ $SU(4)$ irreducible representation.

We will first give the result and later show how this follows from the 
``twisted chiral superfields'' of ref. \cite{hw}.

The lowest order polynomial is the quadratic polynomial in the singletons.
This superfield is the supercurrent multiplet of ref. \cite{hst} and was
 discussed in  \cite{ff1},\cite{ff2},\cite{ffz}.

Because this superfield satisfies some constraints which imply ``current 
conservation'' for spin 2,
spin $\frac{3}{2}$ and spin 1 currents in the ${\bf 1}$, ${\bf 4}$
 (${\bf \bar 4}$) and ${\bf 15}$ reps. of 
$SU(4)$, it is easily seen that it contains $256 = 128_B + 128_F$ states.

The $(E_0,J_1,J_2)$ assignment of these states was given in ref. \cite{ffz}.

Since $256 =2^8$ it then follows that this is a representation of a Clifford algebra with
 8 creation and 8
distruction operators or, equivalently, a superfield with 8 $\theta $'s 
(anticommuting) coordinates.

An $N=$4 chiral superfield would have such property, but its highest spin would
 be a spin 2 in the $(2,0)$ rep. of $SO(4)$.\\
Instead, in order this superfield to contain spin 2 in the graviton
 representation $(1,1)$, one must have what we call
a ``twisted chiral'' superfield, where out of the eight $\theta$'s four are
 2-left-handed spinors
and four are 2-right-handed spinors (instead of four 2-left-handed spinors as
 for a
 $N=4$ chiral superfield).
We call such a superfield twisted chiral.
It is easy to see that the independent components of such a superfield go up
 to $\theta^4\bar \theta^4$.
Moreover it contains a scalar of weight $E_0 =p$ in the $\theta =0$ component
 which is in the $(0,p,0)$ rep. of $SU(4)$,
 a $(1,1)$ rep. of $SO(4)$ with $E_0 = p+2$ in the middle $\theta^2\bar\theta^2
$ component and a real
scalar, in $(0,p-4,0)$ $SU(4)$ rep. in its highest $\theta^4\bar\theta^4$
  component, with $E_0 = p+4$.

This superfield is generic for $p\ge 4$, obtained by taking a $p$-order
 polynomial in the singleton superfield.

It corresponds to the $p\ge 4$ $SU(2,2/4)$ reps. discussed in the previous 
section where the $\theta$-component 
expansion contains $256 \times \frac{1}{12} p^2(p^2-1)$ states.\\
This number is obtained by multiplying the $\theta$ expansion degeneracy with
 the $SU(4)$ rep. of the spin 2 state.
This state lies in the $(0,p-2,0)$ rep. with dimension $\frac{1}{12} p^2(p^2-1)$.

It is obvious that for $p=2,3$ the superfield is not generic because its
 expansion goes up to 
$\theta^2\bar\theta^2$ ($p=2$) and $\theta^3\bar\theta^3$ ($p=3$).
Still the degeneracy is given by the above formula for $p=0,1$.
Note that in these cases the states of highest weights have $E_0= p+2$ and
 $E_0 = p+3$ respectively.
\vskip 5mm

In the second part of this section we recall the concept of ``twisted
 chiral'' superfield of the
$N=4$ four dimensional superconformal algebra \cite{hw}.\\
The singleton superfield (Lie algebra valued in $SU(n)$) is $W_{AB}$ and it
 satisfies the constraints \cite{hst}
\ba
W_{[AB]} &=& \half \epsilon_{ABCD} \bar W^{[CD]}\\
{\cal D}_{\alpha A} W_{[BC]}&=& {\cal D}_{\alpha [A} W_{BC]}. 
\ea
The $(x, \theta_\alpha,\theta_{\dot\alpha})$ superspace can be extended to
 ``harmonic superspace'' \cite{gikos},\cite{hw}
by introducing extra coordinates, elements of the coset $SU(4)/S(U(2)\times
 U(2))$ \cite{hw}.\\
Then the superfield carries an induced representation of $SU(4)$ with isotropy
 group $SU(2)\times SU(2) \times U(1)$.

Let us define $U^A_I$ as an element $U\in SU(4)$.
\be
U^A_I \equiv (u^A_r , u^A_{r'}) \quad r,r'=1,2 \quad(A=1,\cdots ,4)
\ee
Then we can introduce the superfield $W(x,\theta,\bar \theta ,u)$:
\be
W =\half \epsilon^{rs} u^A_r u^B_s W_{[AB]}(x,\theta,\bar \theta)
\label{w}
\ee
It satisfies the following properties \cite{hw}:
\begin{eqnarray}
D_{\alpha r} W &=& D_{\dot\alpha}^{r'} W =0  
\label{grass}\\
D_r^{s'} W&=&0 
\label{analit}\\
{\cal D}_0 W &=&W
\label{wpeso}  
\end{eqnarray}
where:
\begin{eqnarray}
  D_{\alpha I} &=& U^{\ A}_I {\cal D}_{\alpha A},\\
D_{\dot \alpha}^I &=& (U^{-1})^{\ I}_A {\cal D}_{\dot\alpha}^A,
\end{eqnarray}
$D_0$ denotes the derivative with respect to the $U(1)$ part of the isotropy group 
and:
\be
D_I^{\ J} : D_I^{\ J}U_K^{\ B} = \delta _K^J U_I^{\ B}-\frac{1}{4} \delta_I^J
 U_K^{\ B}
\ee
Properties \eq{grass} and \eq{analit} were called Grassmann-- and
 F--analiticity in \cite{hw}.
\\
Since \eq{grass} and \eq{analit} are linear it is obvious that
\begin{equation}
  \label{a}
  A_p = TrW^p
\end{equation}
satisfies \eq{grass} and \eq{analit} together with
\begin{equation}
  {\cal D}_0 W =pW 
\label{apeso} 
\end{equation}
Moreover, eq. \eq{grass} means that $A_p$ depends only on $\theta_{\alpha r'}$
 and 
$\theta_{\dot\alpha}^r$, i.e. on 2 left and 2 right $\theta$'s.\\
It then follows that the superfield \eq{a} has a $\theta$ expansion which only
 contains 
$(J_1,J_2)$ representations with $J_1 \le 1$, $J_2 \le 1$ with highest spin 2
 in the $(1,1)$
representation as discussed in the previous section.

The superfields \eq{a}, called ``primary superfield'' in ref. \cite{hw},
corresponds to the K--K excitations previously discussed.
In particular the singleton composite $W^p$ precisely describes a short
 representation of the $SU(2,2/4)$
algebra discussed in \cite{gm} where a unitary representation is denoted
 by a number $p$
and a given state by the quantum numbers $(E_0, J_1, J_2, (a_1,a_2,a_3))$ where
 $(a_1,a_2,a_3)$
is the Dynkin label of the given $SU(4)$ representation.

It is straightforward to see that the $\theta=0$ component of $A_p$, denoted by
 $\phi_{A_1B_1\cdots A_pB_p}$,
defined by:
\begin{equation}
  Tr\epsilon^{r_1s_1}\cdots\epsilon^{r_p s_p}u_{r_1}^{\ A_1}
 u_{s_1}^{\ B_1}\cdots u_{r_p}^{\ A_p} u_{s_p}^{\ B_p}
\Phi_{A_1B_1\cdots A_pB_p}
\end{equation}
precisely corresponds to a scalar in the $(0,p,0)$ $SU(4)$ representation with
 $E_0=p$.

The $\theta^4\bar\theta^4$ component is a real scalar
\be
 \phi_{A_1B_1\cdots A_pB_p} (\theta^4\bar\theta^4)
=(\theta^4\bar\theta^4)_{\{A_1B_1\cdots A_4B_4}\Phi_{A_5B_5\cdots A_pB_p\}}-
 Traces
\ee
which is in the $(0,p-4,0)$ $SU(4)$ rep. with $E_0 =p+4$.
This state exists for $p\ge 4$.

The K--K graviton excitations lie in the $\theta^2\bar\theta^2$ component, they
 belong to the
$(0,p-2,0)$ $SU(4)$ rep. and have $E_0 =p+2$.

For $p=2$, $Tr W^2$, as shown in ref. \cite{ffz}, is the supercurrent multiplet
 of ref. \cite{hst},\cite{brw}
and describes  the massless graviton multiplet of the Anti-de Sitter bulk, it
 is a bilinear
composite of singletons, as already shown in  \cite{ff2}.

For a given $p$, the total number of component fields is precisely $2^8 \times
 \frac{1}{12}p^2(p^2-1)$
($p\ge 2$).

All superfields with $p>2$ are massive.

By expanding $Tr W^p$ in powers of $\theta$ one finds precisely the massless
 and massive spectrum,
described in ref. \cite{krv}, of type IIB string on $AdS_5 \times S^5$,
 in terms of composite singleton excitations.

The K--K masses are simply obtained by using the formulae of ref. \cite{ffz}:
\begin{itemize}
\item {Bosons:
    \begin{eqnarray}
      m^2_{(0,0)} &=& E_0(E_0 - 4) \nn\\
      m^2_{(1,0)} &=& (E_0 - 2)^2 \nn\\
      m^2_{(\half,\half)} &=& (E_0 - 1)(E_0 - 3) \nn\\
      m^2_{(1,1)} &=& E_0(E_0 - 4) 
    \end{eqnarray}}
\item{Fermions:
    \begin{equation}
      |m_{(\half , 0)}| = |m_{(0,\half )}| = E_0 - 2
    \end{equation}}
\end{itemize}

We note in particular that, as a consequence of superfield multiplication,
we can read out which specific singleton composite corresponds to a given
 state of the K--K
spectrum, as given in \cite{krv}.
As an example we note that the complex scalar $F^2 + {\rm i} F\tilde F$
\footnote{Here $F$ denotes the $SU(n)$ gauge field strength.}
 (the
 axion--dilaton)
corresponds to a $m^2 = 0$ state of the $p=2$ (supercurrent) multiplet
\cite{wit},\cite{ffz}.

A real ``$F^4$'' scalar
\footnote{We benefited of a discussion with Juan Maldacena on this particular
 composite operator.}
 corresponds to a $m^2=32$ state of the $p=4$ massive
 composite \cite{krv}.

No other $F^p$ terms ($p>4$) appear as singleton composites in the K--K tower.

\section{Theories with lower supersymmetries}

For $N=1,2$ four dimensional superconformal theories, corresponding to
 singleton field theories of $SU(2,2/1)$, 
$SU(2,2/2)$ superalgebras, the analysis of K--K excitations can be done in
 terms of a ``chiral''
singleton gauge field strength and additional ``chiral'' and/or
 ``hypermultiplets''.

For $N=2$ superconformal field theories, the $N=4$ $AdS_5$ supergravity has 
three kinds of bulk
multiplets: the graviton, tensor and gauge multiplets.

For $N=1$ superconformal field theories the bulk multiplets are four,
 the graviton, gauge, tensor and hyper multiplet \cite{fz},\cite{grw}.

All these multiplets can be obtained as suitable $N=1$ composites of singleton
 multiplets
of chiral (antichiral) type, i.e. the $N=1$ chiral field strength $W_\alpha$
  and additional chiral matter.

Due to the fact that the matter content in theories with lower supersymmetry
 is model dependent,
the analysis of K--K excitations in terms of boundary composite fields is more
 complex in this case. 

The analysis of the K--K spectrum for a class of these theories has just appeared in the literature \cite{new}.

\section{Conclusions}
In this work we have shown that the towers of K--K excitations of IIB supergravity on $AdS_5 \times S^5$ 
is described by precisely the infinite sequence of ``primary'' analytic conformal $N=4$ superfields 
discussed in \cite{hw}.
There it was shown that their correlation functions in harmonic superspace are holomorphic sections of
a line bundle.
We prefer to call these superfields ``twisted chiral'', in analogy with a
 similar structure
in $N=2$ $2D$ superconformal field theories \cite{2d}.

From our correspondence and from the proposal in \cite{gkp},\cite{wit} it emerges that 
such holomorphic correlators are precisely the boundary values of $N=8$ supergravity amplitudes
in $AdS_5$.
It would be interesting to explore the consequences of this relation in the context of
CFT/AdS duality.

\section*{Acknowledgements}
 We thank J.Maldacena, K.Stelle, L. Castellani and A. Zaffaroni for interesting
discussions.

\end{document}